\documentclass[12pt,a4paper]{article}

\usepackage[dvips]{graphicx}
\usepackage{amsmath}
\newcommand{\be}{\begin{eqnarray} }
\newcommand{\ee}{\end{eqnarray} }
\newcommand{\beq}{\begin{equation} }
\newcommand{\eeq}{\end{equation} }
\newcommand{\e}{\; {\rm e} }
\newcommand{\E}{\; {\rm Exp} }
\def\qq{{\rm Q}_{S}}
\def\q{{\rm Q}_{NS}}
\def\Q{{\hat {\rm Q}}_{NS}}

\def\a{\alpha_s}
\def\A{{\hat\alpha}_s}

\begin{document}
\begin{center}
         {\large \bf 
Direct connection between the different QCD orders for parton distribution 
and fragmentation functions
}
 \end{center}
 \begin{center}
 \vskip 1cm
 {\large O.\,Yu.\,Shevchenko}\footnote{E-mail address: shev@mail.cern.ch}

 \vspace{0.5cm}
 {\it Joint Institute for Nuclear Research}
 \end{center}

\begin{abstract}
The formulas directly connecting
parton distribution functions (PDFs) and fragmentation functions (FFs) at 
the next to leading order (NLO) QCD with the same quantities at the leading 
order (LO) are derived. 
These formulas are universal,  
i.e. have the same form for all kinds of PDFs and FFs, differing only in the 
respective splitting functions entering there.

\end{abstract}
\begin{flushleft}
\end{flushleft}

The extraction of PDFs and FFs from the experimental
data is one of the important tasks of the modern hadron physics.
The most simple and transparent way to do it is the QCD analysis of the data on
measured asymmetries and cross-sections
in LO QCD. 
One of the main advantages of such analysis is that the central values and 
uncertainties of measured asymmetries and cross-sections directly propagate    
to the central values and errors of PDFs and FFs extracted from these
data in LO QCD
(see, for instance, Fig. 3 in \cite{compass2010}).
At the same time the situation with NLO analysis is much more difficult
because instead of simple algebraic equations (see, for example, Eq. (2) in 
Ref. \cite{compass2010}) one deals 
there with complex integral equations (like, for instance, Eqs. (10)-(13) in 
Ref. \cite{EPJ2010}) for finding PDFs (FFs) we are interested in.
The standard
way to solve this problem is to apply the QCD analysis based on the fitting 
procedure (see \cite{EPJ2010} and references therein).   
However, there are unavoidable ambiguities inherent in a fitting procedure 
which become especially important when the quality of the fitted data is rather 
bad (small number of points with large errors). 
These are
arbitrariness in the choice of the functional form (with a lot of varied 
parameters)
of the fitted PDFs and FFs  
at initial scale
and also
ambiguities in the error band calculation (ambiguities related to the deviation 
of $\chi^2$ profile from 
the quadratic parabola and to the choice of $\Delta\chi^2$ determining 
the uncertainty size -- see discussion on this subject in Ref. \cite{EPJ2010}). 
Thus, it seems to be very useful
if one could obtain NLO (NNLO, ...) results on PDFs/FFs using the respective LO 
results as an input, without loosing, thereby, all advantages of LO analysis. 

We start with some necessary notation and definitions. 
For the flavor non-singlet and singlet quantities we introduce the notation $\q$
and ${\bf V}=\left(\qq, {\rm G} \right)$, where
$\q$ can be either $q_{NS}$ (non-singlet combinations of quark densities), or
$\Delta q_{NS}$ (non-singlet combinations of helicity PDFs), 
or combinations of transversity PDFs 
$\Delta_T q (\bar q) \equiv h_{1q,\bar q}$,$\ldots$, 
or $D_{NS}^h$ (``non-singlet'' combination of FFs $D_q^h$),$\ldots$, while 
$\qq$ can be $q_{S}$, $\Delta q_{S}$, $D_{S}^h$,$\ldots$,   
${\rm G}$ can be $g$, $\Delta g$, $D_{g}^h$,$\ldots$.  
In this notation the DGLAP evolution equations (see \cite{Altarelli:1981ax} for 
review) look as
\be
\label{dglap}
Q^2 d{\bf V}(Q^2,x)/dQ^2=
(\a/2\pi)[{\bf P^{(0)}}(x) + (\a/2\pi){\bf P^{(1)}}(x)+ O(\a^2)]\otimes 
{\bf V}(Q^2,x),
\ee
where the convolution ($\otimes$) is given by
\begin{equation}
(A\otimes B)(x)= \int^1_0 dx_1\int^1_0dx_2\, \delta(x-x_1x_2)A(x_1) B(x_2) = \int^1_x \frac{dy}{y} A(\frac{x}{y}) B(y),
\nonumber
\end{equation}
and analogously for $\q$ with the replacement 
${\bf P}(x,\a) \to P(x,\a)= P^{(0)}(x) +(\a/2\pi) P^{(1)}(x) + O(\a^2)$.
Here
${\bf P}$ is $2\times 2$ matrix 
with the elements 
$P_{qq},\, P_{qg},\, P_{gq},\, P_{gg}$, 
and the splitting functions for unpolarized PDFs and helicity PDFs can be found 
in the review \cite{LAMPE},
for transversity PDFs -- in the review \cite{BARONE},
for FFs -- in Ref. \cite{NASON} and references therein. 

Following  \cite{furman} it is convenient  
to define the evolution operators 
 ${\rm E}$ and ${\bf E}$ ($2\times 2$ matrix 
with the elements $E_{qq},\, E_{qg},\, E_{gq},\, E_{gg}$) as
\beq
\label{re} 
\q(Q^2,x) = {\rm E(Q^2,x)}\otimes  \q(Q_0^2,x), \quad  
{\bf V}(Q^2,x) = {\bf E}(Q^2,x)\otimes  {\bf V}(Q_0^2,x).
\eeq
Here we are interested in the initial conditions\footnote{We do not consider 
the asymptotic conditions \cite{furman}
$ {\bf E}\,({\rm E})  \to
 {\bf \hat E}\,(\hat {\rm E})$ as $Q^2\to \infty$ 
(see Eq. (5.57) in \cite{furman}), 
since we deal only with particular 
realization (\ref{re}) of the general conditions given by Eqs. (5.18) in 
\cite{furman}.} 
\beq
\label{bc} 
{\rm E}(Q^2=Q_0^2,x)=\delta(1-x),\quad {\bf E}(Q^2=Q_0^2,x)={\bf 1}\,\delta(1-x),
\eeq
which allow to evolve $\q$ and $ {\bf V}$ from the initial scale $Q_0^2$ to an 
arbitrary scale $Q^2$.

It is also convenient to use, following  \cite{furman}, 
the evolution variable 
$t=(2/\beta_0)\ln\left(\a(Q_0^2)/\a(Q^2)\right)$  
instead of the standard variable $\ln(Q^2/\mu^2)$.
Besides, we introduce the notation 
\be
A \bigl|_{LO}\equiv \hat A, \quad  A \bigl|_{NLO}\equiv A,
\ee
for any quantity $A$ 
at
LO and NLO, respectively.

From now on we consider only the nontrivial singlet case. 
Transition to the simple non-singlet case will be easily done
in the end of calculations by making the replacement of the matrices with the 
respective commuting quantities.  

In terms of quantities $t$ and ${\bf E}$ the DGLAP equations 
are rewritten in LO as
\beq
\frac{d}{d t} {\bf \hat E}(\hat t,x)=
{\bf P^{(0)}}\otimes {\bf \hat E}(\hat t,x),
\label{evolLO}
\eeq
while in NLO they look as
\beq
\label{evolNLO}
\frac{d}{d t} {\bf E}(t,x)=
\left[{\bf P^{(0)}}(x)+ \frac{\a}{2\pi}\,{\bf R}(x)+O(\a^2)\right]\otimes 
{\bf E}(t,x),
\eeq
where 
\beq
\label{r}
{\bf R}(x)\equiv {\bf P^{(1)}}(x)-\frac{\beta_1}{2\beta_0}{\bf P^{(0)}}(x).
\eeq
Solution of (\ref{evolLO}) with the initial condition (\ref{bc})
${\bf \hat E}(\hat t=0,x)={\bf 1}\,\delta(1-x)$
reads \cite{furman}   
\beq
\label{lo}
{\bf \hat E}(\hat t,x)=\E\bigl( {\bf P^{(0)}}(x)\,\hat t \bigl) 
= {\bf 1}\,\delta(1-x) + \hat t\, {\bf P^{(0)}}(x)+ 
\frac{{\hat t}^2}{2!}\,{\bf P^{(0)}}(x) 
\otimes {\bf P^{(0)}}(x) + \ldots,
\eeq
while to solve NLO equation (\ref{evolNLO}) one can apply the elegant method of 
Ref. \cite{furman} based on the
analogy with the perturbative quantum mechanics (see Eqs. (5.47)--(5.54) in 
Ref. \cite{furman}).
Operating in this way one obtains the general solution of (\ref{evolNLO}) in 
the form (for a moment we omit 
$x$ dependence and $\delta(1-x)$)
\be
\label{gen}
{\bf E}(t)=
\left\{ {\bf \hat E}(t)
\otimes \left[{\bf 1} 
+ \frac{\a(Q_0^2)}{2\pi}\int_{t^{'}}^td\tau\,
\e^{-\beta_0\tau/2}\,   
{\bf \hat E}(-\tau)
\otimes {\bf R}
\otimes {\bf \hat E}(\tau)
\right] \otimes 
{\bf \hat E}(-t^{'})\right\} 
\otimes {\bf E}(t^{'}).
\ee

Putting $t^{'}\to \infty$ in (\ref{gen}) one reproduces the solution 
(Eq. (5.54) in Ref. \cite{furman}) 
satisfying the boundary 
condition $ {\bf E} \to {\bf \hat E}$ as $t \to \infty$. 
In turn, putting $t^{'}=0$ in (\ref{gen}) one gets the solution 
\be
\label{our_bc}
{\bf E}(t)= 
 \left[{\bf 1} 
 + \frac{\a(Q^2)}{2\pi}\int_0^t d\tau\,\e^{\beta_0\tau/2}\,   
{\bf \hat E}(\tau)
\otimes {\bf R}
\otimes 
{\bf \hat E}(-\tau)
\right] 
\otimes  
{\bf \hat E}(t),
\ee
satisfying the boundary 
condition (\ref{bc}) we deal with.

The {\it key point} to proceed is the condition that all PDFs and FFs should 
take the same 
values in LO and NLO (as well as in NNLO,$\ldots$) as $Q^2 \to \infty$:
\be
\label{asympt}
\q(Q^2\to \infty,x) = \Q(Q^2\to \infty,x), \quad  
{\bf V}(Q^2\to \infty,x) = {\bf \hat V}(Q^2\to \infty,x).
\ee

Though this asymptotic condition seems to be intuitively clear,
let us argue it in some detail because of its great importance for what follows.

Imagine that two researchers 
analyse in LO (the first) and NLO (the second) the same 
``ideal'' data -- the data available with tremendous statistics even in the 
Bjorken ``sub-limit'' (so high $Q^2$ values
are accessible that the Bjorken scaling violation becomes invisible even within 
extremely small uncertainties
on measured asymmetries and cross-sections). 
For determinacy and simplicity let us suppose that they  
analyse the imaginary ``ideal'' polarized SIDIS data on pion production
and
extract 
the valence helicity PDFs $\Delta u_V$, $\Delta d_V$ from the proton and 
deuteron difference asymmetries 
(see Ref. \cite{shev} and references therein)
measured in the
Bjorken ``sub-limit''.  
The first uses LO formulas 
$A_p^{\pi^+-\pi^-} \sim (4\Delta u_V-\Delta d_V)/(4u_V-d_V)$ and 
$A_d^{\pi^+-\pi^-} \sim (\Delta u_V+\Delta d_V)/(u_V+d_V)$ (i.e., performs 
the analysis analogous to one of COMPASS \cite{compass2008}),
and the second their NLO generalization (Eqs. (6-10) in Ref. \cite{shev}). 
Besides, for self-consistency, both imaginary researches do not use the 
existing parametrizations on $u_V$, $d_V$
but extract these quantities themselves
(as well as the integrated over cut in $z$
difference\footnote{On simultaneous determination
of valence PDFs and $D_1-D_2$ from the SIDIS data see, for example, \cite{EMC}.}
$D_1-D_2$ of favored and unfavored 
pion FFs) 
using the same SIDIS data on pion production averaged over spin and
studying the quantities 
$F_{2p(d,{}^3{\rm He},\ldots)}^{\pi^+}- F_{2p(d,{}^3{\rm He},\ldots)}^{\pi^-} $,                                                    
where in both LO and NLO only  $u_V$, $d_V$ and $D_1-D_2$ survive.
It is obvious that all terms with convolutions $\otimes$  (see Eqs. (6-10) in 
Ref. \cite{shev}) 
distinguishing NLO and LO equations for finding $\Delta u_V$, $\Delta d_V$ and  
$u_V$, $d_V$, $D_1-D_2$ just disappear as one approaches the Bjorken limit, so 
that comparing the results
on these quantities obtained in the Bjorken ``sub-limit''  both researchers 
could not discriminate between them.

So, let us pass to limit $Q_0^2\to \infty$ in Eq. (\ref{re}) using the 
asymptotic condition (\ref{asympt}).  
Then, on the one hand (NLO evolution)
\be
\label{1}
{\bf V}(Q^2,x) ={\bf E}(t\to -\infty,x)\otimes  {\bf V}(Q_0^2\to \infty,x)=
{\bf E}(t\to -\infty,x)\otimes {\bf \hat V}(Q_0^2\to \infty,x), 
\ee
and, on the other hand (inverse LO evolution)
\be
\label{2} 
{\bf \hat V}(Q_0^2\to \infty,x)= {\bf \hat E}(\hat t\to \infty,x)\otimes 
{\bf \hat V}(Q^2,x).
\ee
Combining Eqs. (\ref{1}) and (\ref{2}) one obtains
\be
{\bf V}(Q^2,x)= \left[ \lim_{Q_0^2 \to \infty} {\bf E}(t,x) \otimes 
{\bf \hat E}(-\hat t,x) \right] \otimes
{\bf \hat V}(Q^2,x).
\ee
Using Eqs. (\ref{lo}), (\ref{our_bc}) and 
the relation $\lim_{Q^2 \to \infty} (\a/\A )=1$
we
arrive at the connection formula between NLO and LO 
flavour singlet PDFs (FFs) ${\bf V}$ and ${\bf \hat V}$ at the same finite 
$Q^2$ value
\be
\label{main}
&{\bf V}(Q^2,x)=
\left[{\bf 1}\,\delta(1-x) - \frac{\a(Q^2)}{2\pi}\int_{-\infty}^0 d\tau\,
\e^{\beta_0\tau/2}\,   
{\bf \hat E}(\tau,x)
\otimes {\bf R}(x)
\otimes 
{\bf \hat E}(-\tau,x)
\right] 
\nonumber \\
&\otimes  
\E \left(-\frac{2}{\beta_0}\ln\frac{\a(Q^2)}{\A(Q^2)}\, {\bf P^{(0)}}(x) \right)
\otimes {\bf \hat V}(Q^2,x),
\ee
where 
all dependence on the unreachable infinite point $Q_0^2$ just cancels out.

In the non-singlet case the relation (\ref{main}) is significantly simplified. 
The terms  $\hat E(\tau,x)
\equiv \E(\tau P^{(0)}(x))$ and $ \hat E(-\tau,x)$ cancel out each other in the 
integrand and one
easily obtains
\be
\label{main_NS}
&\q(Q^2,x)=    
\left[\delta(1-x)+ \frac{\a(Q^2)}{2\pi}\left(\frac{\beta_1}{\beta_0^2} 
P^{(0)}(x) -\frac{2}{\beta_0} P^{(1)}(x) 
\right) \right] \nonumber \\ 
&\otimes
\E \left(-\frac{2}{\beta_0}\ln\frac{\a(Q^2)}{\A(Q^2)}\, P^{(0)}(x) \right) 
\otimes
\Q(Q^2,x).
\ee

Eqs. (\ref{main}) and (\ref{main_NS}) connecting flavour singlet and 
non-singlet quantities in NLO with the 
same quantities in LO
is the {\it main result} of the paper. Let us briefly discuss their
practical use. 

There are not any problems with application of Eq. (\ref{main_NS}) and the task 
of reconstruction of NLO non-singlet 
quantities
from LO ones is reduced just to the trivial calculation of the integrals 
entering the convolutions $\otimes$. 
Indeed, the parameter $\epsilon \equiv
-(2/\beta_0)\ln(\a/\A)$
is very small even at the minimal (the lower boundary of the experimental cut on 
$Q^2$ is usually about $1\,GeV^2$) 
really available $Q^2$ values, so that  
one can achieve very good accuracy keeping only few first terms in 
the expansion 
$
\E \left (\epsilon\, P^{(0)}(x)\right )
= \delta(1-x) + \epsilon\,  P^{(0)}(x)+(\epsilon^2/ 2!)\,P^{(0)}(x) 
\otimes P^{(0)}(x) + \ldots 
$
Certainly, the same statement holds for term 
$\E \left (\epsilon\, {\bf P^{(0)}}(x)\right)$ in Eq. (\ref{main}),
but there arises an additional problem 
how to deal with the  integral over $\tau$. As usual, the problem is easily 
solved in the space of Mellin moments.
Notice that $Q^2$ independent integral over $\tau$ in Eq. (\ref{main}) just 
coincides\footnote{
Using Eq. (5.28) in Ref. \cite{furman} for $U$ and  
the obvious relation 
$Q^2d \bigl[\E \left((2/\beta_0) \ln\A \, {\bf P^{(0)}} \right)
\otimes {\bf \hat V} \bigl]/dQ^2=0$ 
one can immediately check
that r.h.s. of Eq. (\ref{main}) indeed satisfies
the NLO DGLAP equation (\ref{dglap}). 
} with the quantity $-U(x)$
in Ref. \cite{furman} (see Eq. (5.45) in \cite{furman}), which enters the 
solution of DGLAP with the 
boundary conditions  
$ \lim_{Q^2\to \infty}{\bf E}\,({\rm E})={\bf \hat E}\,(\hat {\rm E})$ 
(see footnote 2). 
Then, applying the inverse Mellin transformation, one easily obtains instead of 
(\ref{main})
the formula suitable\footnote{In Ref. \cite{vogt} one can find the	
efficient algorithm for the numerical calculation of the integral over $n$ 
(proper choice of the integration contour, etc. --
see discussion around Eq. (3.2) in Ref.  \cite{vogt}).}   
for numerical calculations
\beq
\label{main_n}
{\bf V}(Q^2,x)=
\left[{\bf 1}\delta(1-x)+ \frac{\a(Q^2)}{2\pi}
\int_{C-i\infty}^{C+i\infty} 
dn \frac{x^{-n}}{2\pi i} U(n)   
\right] 
\otimes
\E\left(\epsilon(Q^2) {\bf P^{(0)}}(x)\right) 
\otimes {\bf \hat V}(Q^2,x),
\eeq
where $2\times2$ matrix $U(n)\equiv \int_{0}^{1}\,dxx^{n-1}\,U(x)$ is given by 
Eq. (5.41) in Ref. \cite{furman}.

{\it In summary},
the formulas allowing 
to transform LO parton distribution and fragmentation functions to NLO ones are 
derived.
To obtain these formulas 
we use as an input only the DGLAP evolution equations and the asymptotic 
condition 
that PDFs (FFs) at different QCD orders become the same in the Bjorken limit.
Due to universality of this input the connection formulas are also universal,  
i.e. they are valid for any kind of PDFs (FFs) we deal with.
Besides, it is obvious that operating in the same way 
one can also establish the connection of 
PDFs (FFs) at LO (as well as at NLO) with these quantities at any higher 
QCD order (NNLO, NNNLO, $\ldots$), 
and the only restriction here 
is the knowledge   
of the respective splitting functions.


\begin{thebibliography}{99}

\bibitem{compass2010}
  M.~Alekseev {\it et al.}  [COMPASS Collaboration],
  Phys.\ Lett.\  B {\bf 693} (2010) 227.

\bibitem{EPJ2010} 
  A.~N.~Sissakian, O.~Y.~Shevchenko and O.~N.~Ivanov,
  Eur.\ Phys.\ J.\ C {\bf 65} (2010) 413.

\bibitem{Altarelli:1981ax}
  G.~Altarelli,
  Phys.\ Rept.\  {\bf 81} (1982) 1.

\bibitem{LAMPE}
  B.~Lampe and E.~Reya,
  Phys.\ Rept.\  {\bf 332} (2000) 1.
  
\bibitem{BARONE}
  V.~Barone, A.~Drago and P.~G.~Ratcliffe,
  Phys.\ Rept.\  {\bf 359} (2002) 1.

\bibitem{NASON}
  P.~Nason and B.~R.~Webber,
  Nucl.\ Phys.\ B {\bf 421} (1994) 473.

\bibitem{furman}
 V.~Furmanski and R.~Petronzio,                                   
 Z.\ Phys.\ C {\bf 11} (1982) 293.

\bibitem{shev}
  O.~Y.~Shevchenko, R.~R.~Akhunzyanov and V.~Y.~Lavrentyev,
  Eur.\ Phys.\ J.\ C {\bf 71} (2011) 1713.

\bibitem{compass2008}
  M.~Alekseev {\it et al.}  [COMPASS Collaboration],
  Phys.\ Lett.\ B {\bf 660} (2008) 458.

\bibitem{EMC}
  M.~Arneodo {\it et al.}  [European Muon Collaboration],
  Nucl.\ Phys.\ B {\bf 321} (1989) 541.

\bibitem{vogt}  A.~Vogt, Comput.\ Phys.\ Commun.\ {\bf 170} (2005) 65
 (hep-ph/0408244).


\end{thebibliography}
\end{document}